\documentclass[fleqn, letter]{jpsj3}
\usepackage{txfonts}
\title{Spin Conductivity in Two-Dimensional Non-Collinear Antiferromagnets}
\author{Yurika Kubo\thanks{E-mail: kubo@kh.phys.waseda.ac.jp} and \name{Susumu Kurihara}}
\inst{Department of Physics, Waseda University, 3-4-1 Okubo, Shinjuku, Tokyo 169-8555, Japan} 
\abst{We propose a method to derive the spin current operator for non-collinear Heisenberg antiferromagnets. We show that the spin conductivity calculated by the spectral representation with the spin current satisfies the f-sum rule. We also study the spin conductivity at $T=0$ within spin wave theory. We show how the spin conductivity depends on the external magnetic field with changing magnon spectrum. We also find that the spin Drude weight vanishes for any external magnetic field at $T=0$. }

\kword{spin current, antiferromagnet, f-sum rule, Drude weight, linear response theory}

\begin{document}
\maketitle

Spin currents have attracted considerable interest with the development of spintronics in recent years. Spin conductivity is well-studied theoretically in one-dimensional antiferromagnets by many methods including exact diagonalization\cite{PhysRevB.71.184415,PhysRevB.82.104424}. It is also studied in two-dimensional antiferromagnets \cite{PhysRevB.75.214403,PhysRevB.79.064401,chen2013spin}. These theories on spin currents, however, are rather restricted to collinear antiferromagnets. As far as the authors are aware, there seems to be no clear definition of the spin current operator in the case of non-collinear antiferromagnets for which magnetization is generally not conserved.  

One of our main purposes is to introduce a definition of the spin current operator which fulfills the f-sum rule for non-collinear antiferromagnets; typical examples include Heisenberg antiferromagnets in square and triangular lattices under static homogeneous magnetic fields\cite{PhysRevB.82.144402,PhysRevB.57.5013,PhysRevB.79.144416,kawamura1984spin}. Spins cant on each sublattice in magnetic fields, but the spin current operator can still be defined by using the continuity equation as we show explicitly in this paper. 

This Letter is composed as follows. First, we show a way to define the spin current operator for non-collinear quantum antiferromagnets. We consider $S=1/2$ Heisenberg spins on square and triangular lattices to illustrate our method which, we believe, is applicable to much wider classes of antiferromagnets. We move from a laboratory frame to a rotating frame for convenience\cite{PhysRevB.82.144402,PhysRevB.57.5013,PhysRevB.79.144416}. We then introduce a Holstein-Primakoff boson on the rotating frame common to all sublattices, and diagonalize the harmonic part of the Holstein-Primakoff expansion by Bogoliubov transformation. We employ the linear response theory in spectral representation to calculate the spin conductivity, which is found to satisfy the f-sum rule. We find that the Drude weight vanishes. We also show how the magnon excitation spectrum affects the frequency dependence of the spin conductivity when the external magnetic field is varied.

We now propose a method to define spin wave operators that is valid even for non-collinear antiferromagnets. First, we introduce a useful spin wave formalism of non-collinear Heisenberg antiferromagnets on square and triangular lattices. We suppose each spin on different sublattices to be on the $x_0-z_0$ plane of the laboratory frame $(x_0, y_0, z_0)$, and transform it to the rotating frame $(x, y, z)$ for the square lattice following Ref. \citen{PhysRevB.82.144402}
\begin{align}
\begin{split}
{S}^{x_0}_j&= {\rm e}^{{\rm i}{\bf Q} \cdot {\bf r}_j}S_j^z\cos \theta + S_j^x\sin \theta \text{,}\hspace{10mm}
{S}^{y_0}_j=S_j^y\text{,}\\
{S}^{z_0}_j&=S_j^z \sin \theta - {\rm e}^{{\rm i}{\bf Q} \cdot {\bf r}_j} S_j^x \cos \theta \text{,}
\end{split}
\end{align}
where $\bf{Q}=(\pi,\pi)$, and for the triangular lattice\cite{PhysRevB.79.144416}
\begin{align}
\begin{split}
{S}^{x_0}_j&=S_j^z \sin \theta_j + S_j^x\cos \theta_j \text{,}\hspace{10mm}
{S}^{y_0}_j=S_j^y\text{,}\\
{S}^{z_0}_j&=S_j^z \cos \theta_j - S_j^x\sin \theta_j \text{.}
\end{split}
\end{align}
Here, $\theta$ and $\theta_j$ are canting angles to be given below. We stress that spins on each sublattice are now expressed by a simple set of rotated spin operators $S_i^{\mu}(\mu=x,y,z)$ common to all sublattices. 

The model spin Hamiltonian for both lattices are written in the same form in the laboratory frame 
\begin{align}
\hat{H}=J\sum_{<i,j>}\left( S^{x_0}_iS^{x_0}_j+S^{y_0}_iS^{y_0}_j+S^{z_0}_iS^{z_0}_j\right)-h\sum_iS_i^{z_0} \text{.}
\end{align} Here, $J$ denotes the exchange constant and $h$ denotes a uniform magnetic field. Magnetization saturates at $h= 8JS$ in the square lattice, and spins select up-up-down phase at $h= 3JS$ in the triangular lattice.

Canting angles are determined to minimize the ground state energy. For the square lattice, $\theta$ is given by \cite{PhysRevB.82.144402,PhysRevB.57.5013} 
\begin{align}
\theta=\sin^{-1} \left( \frac{h}{8JS} \right) \text{.}
\end{align}
For the triangular lattice, they are given by\cite{kawamura1984spin}
\begin{align}
\begin{split}
\theta_A&= -\pi \text{,}\hspace{10mm} \theta_B=-\theta_C= \cos^{-1}\left [ \frac{1}{2}\left (\frac{h}{3JS}+1 \right) \right] \text{,}
\end{split}
\end{align}
where $\theta_i$ ($i=A,B,C$) denotes the canting angle for each sublattice. We perform Holstein-Primakoff transformations to spin operators on the rotating  frame with bosons $a_j$: \cite{PhysRevB.82.144402,PhysRevB.57.5013,PhysRevB.79.144416}
\begin{align}
\begin{split}
S_j^+ &=   \sqrt{2S-a^{\dagger}_j a_j}\hspace{1.4mm} a_j \text{,}\hspace{10mm}
S_j^- =a^{\dagger}_j \hspace{0.2mm} \sqrt{2S-a^{\dagger}_j a_j}\text{,} \hspace{0.8mm}\\
S_j^z &= S-a^{\dagger}_j a_j \text{.}
\end{split}
\end{align}
Next, we perform Fourier transformation, and then Bogoliubov transformation with new bosons $b_{\bf k}$:
\begin{align}
a^{\dagger}_k&=u_k b^{\dagger}_{\bf k}+v_k b_{\bf -k} \hspace{10mm} (u_k^2-v_k^2=1)
\end{align}
to diagonalize the harmonic part of the bosonic Hamiltonian.\cite{PhysRevB.82.144402,PhysRevB.57.5013,PhysRevB.79.144416}
In this way, we obtain the spin-wave spectrum and $u^2_k,v^2_k, u_kv_k$ for the square lattice: \cite{PhysRevB.82.144402,PhysRevB.57.5013}
\begin{align}
\begin{split}
\omega_{k}&=4JS \sqrt{ (1+\gamma_{k}) (1- \gamma_{k}\cos 2\theta ) }\text{,}\\
\gamma_k&=\frac{1}{2}(\cos k_x +\cos k_y )\text{,}
\label{omegasq}
\end{split}
\end{align}
\begin{align}
\begin{split}
u_k^2,v_k^2&=\frac{1}{2}(\frac{A_k}{\omega_k}\pm1) \text{,}\hspace{17mm}
u_kv_k=\frac{B_k}{2\omega_k}\text{,}\\
A_k &=4JS(1+\gamma_k \sin^2 \theta ) \text{,} \hspace{10mm}B_k =4JS\cos^2 \theta \text{.}
\label{ukvk} 
\end{split}
\end{align} 
The spin-wave spectrum for the triangular lattice \cite{PhysRevB.79.144416} is
\begin{align}
\begin{split}
\omega_{k}&=3JS \sqrt{ (1+2\gamma_{k}) \left(1+\gamma_{k} \left(\frac{1}{3} \left(\frac{h}{3JS} \right) ^2 -1 \right) \right) }\text{,}\\
\gamma_k&=\frac{1}{3}\left(\cos k_x +2\cos \frac{k_x}{2}\cos \frac{\sqrt{3}k_y}{2} \right)\text{.}
\label{omegatri}
\end{split}
\end{align}
We also get $u^2_k,v^2_k$ and $u_kv_k$ in the same way as in Refs. \citen{PhysRevB.82.144402,PhysRevB.57.5013,PhysRevB.79.144416} for the triangular lattice. We suppress the expressions for $u_k$ and $v_k$ to save space. We see that the energy gap $\Delta=h$ opens at $\Gamma$ point in both lattices when an external magnetic field $h$ is applied.

In view of the apparent lack of a suitable definition of the spin current operator for non-collinear systems, we now focus on the derivation of the operator on the basis of the 
continuity equation. The underlying conservation law of the total magnetization 
$M_{\rm tot}$ is clear in the spin Hamiltonian formalism. This, however, is no longer obvious after we move to Holstein-Primakov boson representation. We must thus examine 
to what extent the conservation law holds in the truncated boson representation, before using the continuity equation.

The bosonic Hamiltonian\cite{PhysRevB.82.144402} $\hat{H}_{\rm b}$ and the total magnetization $M_{\rm tot}=\sum_iS_i^{z_0}$ are given as follows:
\begin{multline}
\hat{H}_{\rm b}=-2JS^2N\cos 2\theta+h \sin \theta\sum_i n_i\\
+JS\sum_{<i,j>}\left[\sin^2 \theta(a^{\dagger}_ia_j+a_ia^{\dagger}_j)-\cos^2 \theta(a_ia_j+a^{\dagger}_ia^{\dagger}_j)\right]\\
+JS\cos 2\theta\sum_{<i,j>}(n_i+n_j)+\cdots\text{,}
\label{bosonhamiltonian}
\end{multline}
\begin{multline}
M_{\rm tot}=SN\sin \theta-\sqrt{\frac{S}{2}}\cos \theta\sum_i {\rm e}^{\rm{i}\bf{Q}\cdot \bf{r}_i}(a_i+a^{\dagger}_i)\\
-\sin \theta\sum_i n_i+\cdots\text{,}
\label{mtot}
\end{multline}
where $n_i=a^{\dagger}_ia_i$. 
We see that the commutator $[H_{\rm b}, M_{\rm tot}]$ is formally a power series in $1/\sqrt{S}$, whose $n$th term is of the order $S^{3-(n-1)/2}$.  We can show that the first six terms vanish.  In other words, the total magnetization operator in boson representation is in fact an approximately conserved quantity up to this accuracy. We are thus justified in the use of the continuity equation to calculate the current operator $j_{{\rm s}\hspace{1mm} i,i+\hat{x}}$. 

Now, we derive the spin current density operators $j_{{\rm s}\hspace{1mm}i,i+\hat{x}}$ for both lattices by spin wave operators we introduced, where $\hat{x}$ is the unit lattice vector in the $x-$direction. The spin conductivity $\sigma_{\rm s}(\omega)$ is defined by the linear response relation
\begin{align}
{\cal J}_{\rm s}&=\sigma_{\rm s}\nabla_{x} h\text{,}
\end{align}
where $\nabla_x h$ is the gradient of magnetic fields $h$ along the $x-$direction and ${\cal J}_{\rm s}$ is the induced spin current. \cite{PhysRevB.71.184415,PhysRevB.82.104424,PhysRevB.75.214403,PhysRevB.79.064401,chen2013spin}
We assume that spin current flows along the field gradient. We now use the continuity equation in the long-wavelength limit, which is written by the local magnetization density $S_i^{z_0}/\Omega$ 
\begin{align}
\frac{1}{\Omega} \partial_t S_i^{z_0} &=-\frac{j_{{\rm s}\hspace{1mm}i,i+\hat{x}}-j_{{\rm s}\hspace{1mm}i,i-\hat{x}}}{a_0} \text{,}
\end{align}
where  $\Omega$ denotes the area of the unit cell and  $a_0$ denotes the lattice constant.
We obtain a spin current density operator by Heisenberg equation of motion $\partial_t S_i^{z_0}={\rm i}\left[\hat{H},S_i^{z_0}\right]$ and $\hat{H}_{i,i+\hat{x}}=J{\bf S}_{i}\cdot{\bf S}_{i+\hat{x}}$:
\begin{align} 
j_{{\rm s}\hspace{1mm}i,i+\hat{x}}&={\rm i}\frac{a_0}{\Omega} \left[\hat{H}_{i, i+\hat{x}},S^{z_0}_{i+\hat{x}}\right]
=\frac{a_0}{\Omega}J\left(S^{x_0}_iS^{y_0}_{i+\hat{x}}-S^{y_0}_iS^{x_0}_{i+\hat{x}}\right) \text{.}
\end{align}
Then, we move from the laboratory frame to the rotating frame in each lattice and perform Holstein-Primakoff expansion. We get the spin current density operator
\begin{align} 
j_{{\rm  s}\hspace{1mm} i,i+\hat{x}}&=\sum_n j_{{\rm s} \hspace{1mm}i,i+\hat{x}  \hspace{1mm}n/2} \hspace{5mm} n=3,2,1 \cdots \text{,}
\end{align} 
where $j_{{\rm s} \hspace{1mm} i,i+\hat{x}\hspace{1mm}n/2}$ is a term proportional to $S^{n/2} $. For the square lattice, the leading term $j_{{\rm s} \hspace{1mm}i,i+\hat{x}\hspace{1mm} 3/2}$ is given by
\begin{multline}
j_{{\rm s} \hspace{1mm} i,i+\hat{x}\hspace{1mm} 3/2}=-\frac{{\rm i}a_0JS}{\Omega}\sqrt{\frac{S}{2}}
 {\rm e}^{{\rm i}{\bf Q}\cdot {\bf r}_i} \cos \theta \left(a_i-a^{\dagger}_i\right)\\
-\frac{{\rm i}a_0JS}{\Omega}\sqrt{\frac{S}{2}
{\rm e} }^{{\rm i}{\bf Q}\cdot {\bf r}_i} \cos \theta\left(a_{i+\hat{x}}-a^{\dagger}_{i+\hat{x}}\right)\text{,}
\label{currentsq}
\end{multline} 
and for the triangular lattice, it is given by
\begin{multline} 
j_{{\rm s} \hspace{1mm}i,i+\hat{x}\hspace{1mm}3/2}
=\frac{{\rm i}a_0JS}{\Omega}\sqrt{\frac{S}{2}}
\sin \theta_{i+\hat{x}}\left(a_i-a^{\dagger}_i\right)\\
-\frac{{\rm i}a_0JS}{\Omega}\sqrt{\frac{S}{2}}
\sin \theta_{i}\left(a_{i+\hat{x}}-a^{\dagger}_{i+\hat{x}}\right) \text{.}
\label{currenttri}
\end{multline}
Here, we use a Holstein-Primakoff boson for simplicity, though we need to perform Bogoliubov transformation to calculate the spin conductivity. We see that the spin current density operators $j_{{\rm s} \hspace{1mm} i,i+\hat{x} \hspace{1mm}3/2}$ are of the first-order in bosonic operators.

Next, we show how we calculate the spin conductivity $\sigma_{\rm s}(\omega)$, which is written by the Drude weight $D_{\rm s}$ and the regular part $\sigma_{\rm s, reg}(\omega)$: 
\begin{align} 
\sigma_{\rm s}(\omega)&=D_{\rm s}\delta(\omega)+\sigma_{\rm s, reg}(\omega)\text{.}
\end{align}
We refer to Kubo formula for the electrical conductivity\cite{mahan}
and spin conductivity\cite{PhysRevB.82.104424,PhysRevB.75.214403} obtaining the regular part of the spin conductivity $\sigma_{\rm s, reg}(\omega)$ for $T=0$ in spectral representation:\cite{PhysRevB.16.2437}
\begin{align} 
\sigma_{\rm s,reg }(\omega)&=\frac{\pi \Omega}{N}\sum_{E_m\neq E_0} |\left<m|J_{\rm s}({\bf q})|0 \right>|^2\frac{\delta \left(|\omega|-\left(E_m-E_0\right)\right)}{E_m-E_0},
\label{regular}
\end{align}
where $N$ denotes the number of lattice sites, and $J_{\rm s}({\bf q})$ denotes Fourier representation of $j_{{\rm s}\hspace{1mm} i,i+\hat{x}}$. The Drude weight can be evaluated as follows:
\begin{align} 
D_{\rm s}&=\frac{\pi a_0}{N \Omega} \left<-\hat{T} \right>-I_{\rm reg} \label{drude}\text{,}\\
I_{\rm reg}&=\int_{-\infty}^{\infty} \sigma_{\rm s,reg }(\omega){\rm d} \omega  \label{Ireg}\text{.}
\end{align}
The first term in Eq. (\ref{drude}) is related to the f-sum rule for the spin conductivity, which we now discuss in detail.

We derive the f-sum rule for a spin current density operator in this model following Refs. \citen{PhysRevB.75.214403} and \citen{PhysRevB.16.2437}. 
First, we show the continuity equation of both lattices in Fourier representations in the long-wavelength limit:\cite{PhysRevB.75.214403}
\begin{align}
{\rm i} q_x J_{\rm s}({\bf -q})&=\frac{1}{\Omega }\partial_{t}S^{z_0}({\bf -q})\text{,}
\label{fcon}
\end{align}
where $S^{z_0}({\bf q})$ denotes Fourier representation of $S^{z_0}_i$.
Then we calculate the frequency integral of Eq. (\ref{regular}) with the use of Eq. (\ref{fcon})
\begin{align}
 {\rm i}q_x \frac{ N}{\pi }\int_{-\infty}^{\infty} \sigma_{\rm s}(\omega){\rm d} \omega 
=\sum_{m}\frac{\left<0|J_{\rm s}({\bf q})|m \right>\left<m \left|\partial_t S^{z_0}(-{\bf q}) \right|0\right>}{E_m-E_0}\notag \\
-\sum_{m}\frac{\left<0 \left|\partial_t S^{z_0}(-{\bf q}) \right|m \right>\left<m|J_{\rm s}({\bf q})|0 \right>}{-\left(E_m-E_0 \right)}\text{.}
\end{align}
We obtain the following equations by applying Heisenberg equation of motion
\begin{align}
{\rm i}q_x \frac{N}{\pi }\int_{-\infty}^{\infty} \sigma_{\rm s}( \omega){\rm d} \omega 
&=\left<0 \left|{\rm i}\left[J_{\rm s}({\bf q}),S^{z_0}(-{\bf q})\right]\right|0\right>\text{.}
\label{totalweight}
\end{align}
We obtain the left-hand side of Eq. (\ref{totalweight}) by calculating ${\rm i}\left[J_{\rm s}({\bf q}),S^{z_0}(-{\bf q})\right]$ without any approximations using the laboratory frame:
\begin{align}
{\rm i}\left[J_{\rm s}({\bf q}),S^{z_0}({\bf -q})\right] &=-{\rm i}q_x \frac{a_0}{\Omega}\sum_{l}J\left(S^{x_0}_lS^{x_0}_{l+\hat{x}}+S^{y_0}_lS^{y_0}_{l+\hat{x}}\right)\notag \\
 &=-{\rm i}q_x \frac{a_0}{\Omega}\hat{T} \text{.}
\end{align}
Here, $\hat{T}$ denotes $xy$-part of the exchange interaction in the laboratory frame:
\begin{align} 
\hat{T}& = \sum_{l}J\left(S^{x_0}_lS^{x_0}_{l+\hat{x}}+S^{y_0}_lS^{y_0}_{l+\hat{x}} \right)\text{.}
\end{align}
We see that the spin conductivity in both lattices satisfies the f-sum rule by the preceding procedure:
\begin{align} 
\int_{-\infty}^{\infty} \sigma_{\rm s}(\omega){\rm d} \omega &=\frac{\pi a_0}{N \Omega}\left<-\hat{T}\right>
\label{frule} \text{.}
\end{align}
This is the exact form of the spin conductivity f-sum rule valid for any lattice form. 

Next, we examine the left-hand side of Eq. (\ref{frule}), which is denoted as $I$, and classify various terms in $I$ coming from the Holstein-Primakoff expansion according to the powers of $S$:
\begin{align} 
I &=\sum_n I_n \hspace{5mm} n=2,1,0 \cdots \text{,}
\end{align}
where $I_n$ is a term proportional to $S^{n}$.   
We focus on 
\begin{align} 
\left<\hat{T}\right>=\sum_n \left<\hat{T}_{n}\right> \hspace{5mm} n=2,1,0 \cdots \text{,}
\end{align}
where $\hat{T}_n$ is a term proportional to $S^{n}$, to derive the left-hand side of Eq. (\ref{frule}) in Holstein-Primakoff expansion smoothly. We don't have to consider $\hat{T}_{n/2}$ when $n$ is an odd integer, because their expectation values are always zero. 
We calculate $I_n$ using the following equation
\begin{align} 
I_n &= \frac{\pi a_0}{N \Omega}\left< -\hat{T}_n\right> \hspace{5mm} n=2,1,0 \cdots  \text{.}
\label{frule2}
\end{align}

We now examine $\hat{T}_2$ and $\hat{T}_1$, which are the first and second terms in Holstein-Primakoff expansion, for both lattices. For the square lattice,
\begin{align}
\hat{T}_2&=-N JS^2 \cos^2 \theta \text{,}
\end{align}
\begin{align}
\hat{T}_1=2JS\cos^2 \theta \sum_l n_l -N JS^2 \left(\cos^2 \theta^{'}-\cos^2 \theta \right) \notag\\ 
+\frac{JS}{2}\sum_l \left(\sin^2 \theta +1\right)\left(a^{\dagger}_la_{l+\hat{x}}+a^{\dagger}_{l+\hat{x}}a_{l} \right)  \notag\\
+\frac{JS}{2} \sum_l \left(\sin^2 \theta -1\right)\left(a^{\dagger}_la^{\dagger}_{l+\hat{x}}+a_{l}a_{l+\hat{x}} \right)  \text{,}
\end{align}
where $\cos^2 \theta^{'}$ is obtained by considering quantum correction to the canting angle using Eq. (\ref{ukvk}):\cite{PhysRevB.82.144402,PhysRevB.57.5013}
\begin{align}
\cos^2 \theta^{'}&=1-\sin^2 \theta \left(1+2\frac{w}{S}\right) \text{,}\\
w&=\frac{1}{N}\sum_{\bf k}\left[(1-\gamma_{\bf k})v_k^2-\gamma_{\bf k}u_k v_k \right]\label{wdef}\text{.}
\end{align}
For the triangular lattice
\begin{align}
\hat{T}_2&=JS^2\sum_l\sin \theta_{l+\hat{x}}\sin \theta_l \text{,}
\end{align}
\begin{align}
\hat{T}_1=
\frac{JS}{2}\sum_l \left( \frac{\cos^2 \theta-2\cos \theta}{3} -1 \right) \left(a^{\dagger}_la^{\dagger}_{l+\hat{x}}+a_{l}a_{l+\hat{x}}\right)\notag \\
+\frac{JS}{2} \sum_l \left( \frac{\cos^2 \theta-2\cos \theta}{3} +1 \right) \left(a^{\dagger}_la_{l+\hat{x}}+a^{\dagger}_{l+\hat{x}}a_{l} \right) \notag \\
-NJS^2 \left(\frac{\sin^2 \theta^{'}-\sin^2 \theta}{3}\right)
+JS \frac{2\sin^2 \theta}{3}\sum_l n_l \text{,}
\end{align}
where $\cos^2 \theta^{'}$ is obtained by the same procedure as the square lattice,\cite{PhysRevB.82.144402,PhysRevB.57.5013} 
\begin{align}
\cos^2 \theta^{'} &=\frac{1}{4}\left[ 1-\frac{2h}{3JS} \left(1+\frac{w}{S} \right) +\left(\frac{h}{3JS}\right)^2 \left( 1+2\frac{w}{S} \right)\right]\text{,}
\end{align}
and $w$ is defined in Eq. (\ref{wdef}). 
We get the f-sum rule for the first and second terms in Holstein-Primakoff expansion by substituting $\hat{T}_2$ and $\hat{T}_1$ to Eq. (\ref{frule2}). We expect that this formalism is valid and independent of the lattice structure as long as magnetization is a conserved quantity. 

In Fig. 1(a), we show the integrated intensity $I$ of the spin conductivity, and the corresponding quantity $I_{\rm reg}$ for the regular part at $T=0$, to the leading order in Holstein-Primakoff expansion for the square lattice. Here, $I_{\rm reg\hspace{1mm}2}$ denotes the leading contribution to $I_{\rm reg}$, defined in Eq. (\ref{Ireg}), in Holstein-Primakoff expansions. The leading term of the spin conductivity is calculated by substituting $J_{\rm s \hspace{1mm} 3/2}({\bf q})$, which denotes Fourier representation of $j_{{\rm s} \hspace{1mm}  i,i+\hat{x} \hspace{1mm} 3/2}$, for $J_{\rm s}({\bf q})$ in Eq. (\ref{regular}) for each lattices. In Fig. 1(a), we show the integrated intensities $I_{\rm reg\hspace{1mm}2}$ and $I_2$, defined in Eq. (\ref{frule2}). We intentionally shifted the curve for $I_2$ slightly because the two results overlap completely. We thus find that the Drude weight vanishes for the square lattice at $T=0$ for any magnetic field $h$ by comparing Fig. 1(a) to Eq. (\ref{drude}), because the difference between these two results defines the Drude weight. The vanishing Drude weight at $T=0$ is consistent with Refs. \citen{PhysRevB.75.214403,chen2013spin,PhysRevB.79.064401}. 

\begin{figure}
\includegraphics[width=6.5cm,clip]{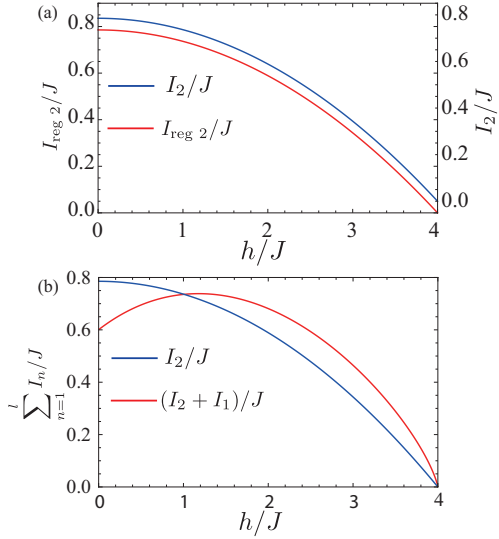}
\caption{(Color) (a) Leading term $I_2$ of the integrated intensity of the spin conductivity and corresponding quantity $I_{\rm reg\hspace{1mm}2}$ for the regular part  as functions of magnetic field on a square lattice. We intentionally shift the curve for figures of $I_2$ because of the degeneracy of the two results, which indicates vanishing of the Drude weight for any field at $T=0$. (b) We compare $I_2$ and $I_2+I_1$ as functions of the magnetic field on the square lattice. $I_2$ monotonically decreases as a result of locking spins with increasing field. Spin wave corrections on staggered magnetization strongly suppress $I_2+I_1$ at low fields.}
\end{figure}

We compare the intensities $I_2$ to $I_2+I_1$ in Fig. 1(b). This figure indicates that there are two kinds of magnetic-field effect on the corrected intensity. One is dominant at low fields and the other is dominant at high fields. Spin wave corrections on staggered magnetization due to zero point fluctuation suppress integrated intensity, and its effect is dominant at low fields with a small gap excitation at $\Gamma$ point. Canting angle changes and saturates with increasing field, locking spins toward the field direction, and thus suppressing the spin conductivity. This effect is dominant at high fields. Whereas the leading term $I_2$ monotonically decreases with the magnetic field $h$, the quantum corrected intensity $I_2+I_1$ is now a non-monotonic function with two kinds of effects. We get similar results in the triangular lattice, which are not shown in this letter.

\begin{figure}
\includegraphics[width=6.5cm,clip]{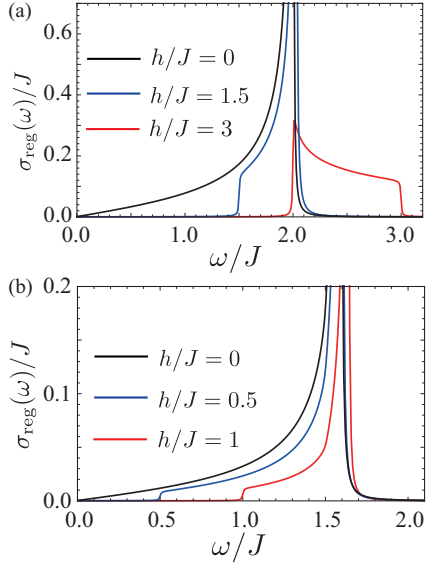}
\caption{(Color) (a) Results of the leading order of spin conductivity for the square lattice. (b) Results of the leading order of spin conductivity for the triangular lattice. These results of both lattices show van-Hove singularities and indicate that the shape of the excitation spectrum affects the results of the spin conductivity.}
\end{figure}

We show the frequency dependence of spin conductivity for the square lattice in Fig. 2(a) and for the triangular lattice in Fig. 2(b), both calculated to the leading order only. We see van-Hove singularity in each lattice for any static homogeneous magnetic field. We expect these singularities to be removed by considering $1/S$ corrections of Holstein-Primakoff expansions \cite{PhysRevB.75.214403}. 

We notice that the lowest and highest values of the excitation spectrum in magnetic Brillouin zone determine the threshold of low- and high-frequency limits of the spin conductivity.  We also note that the lower bound of the spin conductivity spectrum at low fields $h/J\leq1$ is determined by the excitation gap at $\Gamma$ point, which is the lowest energy for both lattices as seen from Eq. (\ref{omegasq}) and Eq. (\ref{omegatri}). On the other hand, the $\Gamma$ point gap determines the upper bound of the conductivity spectrum for sufficiently high magnetic fields, as shown in Fig. 2. We expect a similar behavior in the triangular lattice in higher magnetic field regions. These results indicate that the excitation spectrum of magnetic Brillouin zone essentially determines the spin conductivity. 
\\

In conclusion, we have derived the spin current operator for Heisenberg antiferromagnets and the spin conductivity in square and triangular lattices. We have shown that the Drude weight vanishes at  $T=0$ for any external static magnetic field for a square lattice within the linear spin wave theory, which is consistent with Refs. \citen{PhysRevB.75.214403,chen2013spin,PhysRevB.79.064401}. Two kinds of magnetic-field effect on integrated intensities in spin conductivity are found and the competition between the two makes the corrected intensity a non-monotonic function of the magnetic field $h$. 

We have calculated the frequency dependence of the spin conductivity for both lattices, which indicates that the excitation spectrum of magnetic Brillouin zone determines the spin conductivity. We expect to get more realistic results by taking $1/S$ corrections of Holstein-Primakoff expansions into account. Lastly, we believe that this method is applicable to any value of $S\geqq1/2$ and any non-collinear as well as collinear antiferromagnet as far as magnetization is conserved. 

\bibliographystyle{jpsj}
\bibliography{ref}
\end{document}